\documentclass[preprint2]{aastex}     
\usepackage{xcolor}

\def\kms{{\rm\,km\,s^{-1}}}

\def\Gyr{{\rm\,Gyr}}
\def\deg{{^\circ}}

\def\kpc{{\rm\,kpc}}

\def\mathnew{\mathsurround=0pt}   
\def\simov#1#2{\lower .5pt\vbox{\baselineskip0pt  
    \lineskip-.5pt\ialign{$\mathnew#1\hfil##\hfil$\crcr#2\crcr\sim\crcr}}}

\def\'#1{\ifx#1i{\accent"13\i}\else{\accent"13#1}\fi}

\begin{document}    
\shorttitle{Non-spherical Structure of the Galactic Dark Matter Halo:
  Stellar Kinematics} \shortauthors{Rojas-Ni\~no et al.}

\title{Detecting Triaxiality in the Galactic Dark Matter Halo through
  Stellar Kinematics II: Dependence on Dark Matter and Gravity Nature}

\author{Armando Rojas-Ni\~no$^{1}$, Luis A. Mart\'inez-Medina$^{2}$,
  Barbara Pichardo$^{1}$, Octavio Valenzuela$^{1}$}
 
\affil{$^{1}$Instituto de Astronom\'ia, Universidad Nacional
  Aut\'onoma de M\'exico, A.P. 70-264, 04510, M\'exico, D.F.;
  Universitaria, D.F., M\'exico; \\ $^{2}$Departamento de F\'isica, Centro de Investigaci\'on y de Estudios
Avanzados del IPN, A.P. 14-740, 07000 M\'exico D.F., M\'exico; \\ barbara@astro.unam.mx,octavio@astro.unam.mx}

\begin{abstract} 
Recent studies have presented evidence that the Milky Way global
potential may be non-spherical. In this case, the assembling process
of the Galaxy may have left long lasting stellar halo kinematic
fossils due to the shape of the dark matter halo, potentially
originated by orbital resonances. We further investigate such
possibility, considering now potential models further away from
$\Lambda$CDM halos, like scalar field dark matter halos, MOND, and
including several other factors that may mimic the emergence and
permanence of kinematic groups, such as, a spherical and triaxial halo
with an embedded disk potential. We find that regardless of the
density profile (DM nature), kinematic groups only appear in the
presence of a triaxial halo potential. For the case of a MOND like
gravity theory no kinematic structure is present. We conclude that the
detection of these kinematic stellar groups could confirm the
predicted triaxiality of dark halos in cosmological galaxy formation
scenarios.
\end{abstract}

\section{Introduction}
\label{sec:intro}

The $\Lambda$CDM scenario is considered the standard one in cosmology
because it incorporates self-consistently many large scales properties
of the universe.  An intrinsic feature of this scenario is the
hypothesis of the existence of cold dark matter particles required to
explain the cosmic structure formation. The $\Lambda$CDM model is also
able to make predictions on galactic scales, relating the process of
galaxy formation with the internal properties of galaxies, triggering
a vivid debate in the astronomical
community. \citep[e.g.][]{WReese78,Kauffmann1993,Moore94, Bosch1998,
  Ghigna1998, Klypin1999, Moore99, avila-reese06, Pizagno2007,
  Valenzuela2007,Governato12}.

Cosmological simulations had shown that the formation of galactic size
dark matter halos occur through the hierarchical assembly of CDM
structures \citep{Kauffmann1993,Klypin1999,Moore99}, where small
systems form first and then merge together to form more massive
ones. This process leads to halos with a strongly triaxial shape
\citep{Allgood2006, Vera2011}.

Therefore, the possible detection of the dark matter halos triaxiality 
would be of great importance for these galactic evolution and formation 
theories, as it would confirm one of their most intrinsic predictions.

In paper I \citep{Rojas2012}, we introduced a new strategy in order to
detect triaxiality in the Milky Way halo, in addition to the ones
commonly discussed in literature (\citealt{Jing2002},
\citealt{Law2009}, \citealt{Penarrubia2009}, \citealt{Gnedin2005},
\citealt{Zentner2005}, \citealt{Steffen08},
\citealt{2012ApJ...758L..23L}, \citealt{2012MNRAS.419.1951V},
\citealt{2013MNRAS.434.2971D}, \citealt{Valenzuela14},
\citealt{VeraCiro2014}, \citealt{Deg2013}, \citealt{Lux2012}). In a
more recent paper \citet{2014ApJ...794..151L} solve the
Jeans equation to constraint the shape and distribution of the dark
matter halo within the Milky Way and find out that the solution is
consistent with an oblate halo. The former method is very promising,
although it has an important dependence on assumptions currently
validated mostly by cosmological MW formation simulations which are
still uncertain. Our work is independent and complementary to all of
them. Our strategy is based on the stellar kinematics in the Galaxy
halo. The idea was to compare the orbital structure generated by a
spherical halo with the orbital structure generated with a triaxial
halo.  The kinematic differences between both cases and their
comparison with observational data could help to determine whether the
Galactic halo is really triaxial. An important difference between a
triaxial potential and a spherical one is that in the first case there
are abundant resonant orbits, while in the second rosette orbits are
dominant and have no angular preference.  The presence of resonant
orbits favour the formation of moving groups of stars: the kinematic
groups.

In paper I we showed that if the dark matter halo of the Milky Way
possesses a triaxial shape, the assembling history of the stellar halo
is able to populate the quasi-resonant orbits triggering kinematic
stellar groups.

The analysis of the resulting kinematic and orbital structure may give
evidence for the halo triaxiality.

For the study presented in paper I we only used the gravitational
potential of a NFW dark matter halo with a density profile that
results from CDM-only simulations \citep{Navarro1996}.  But regardless
of the great success of the $\Lambda$CDM scenario explaining the large
scale structure of the universe, the lack of clear positive results in
attempts of direct dark matter detection have motivated the
exploration of alternative models. In this way, with the purpose of
testing other theories, alternative to cold dark matter and their
effect on the dynamical structure of the Galaxy, we introduce two
different models for the potential, one based on modified gravity
theories (MOND specifically), and a different halo model motivated on
Scalar Field Dark Matter (SFDM).

Regarding to MOND, the work of \citet{M1983} is one of the most
popular proposals in astrophysics. A common feature in these family of
theories is that the only gravity source are baryons. In the Milky Way
case, that implies that mostly, the flat disk potential is responsible
of the whole galactic dynamics, particularly in the stellar halo.

On the other hand, regarding the SFDM, this is an alternative scenario
that has received much attention recently, the main hypothesis is that
the nature of the dark matter is determined by a fundamental scalar
field that condensates forming Bose-Einstein Condensate (BEC)
``drops'' \citep{Guzman2000,Magana12}, these condensates represent
galaxies dark matter haloes. \citet{Robles2013} consider that dark
matter is a self-interacting real scalar field embedded in a thermal
bath at temperature $T$ with an initial $Z_2$ symmetric potential. Due
to the expansion of the Universe, the temperature drops in such a way,
that the $Z_2$ symmetry is spontaneously broken, and the field rolls
down to a new minimum.

To study the galactic scale of the model one can solve the Newtonian
limit of the equation that describes a scalar field perturbation,
i.e., where the field is near the minimum of the potential that
describes its interaction and where the gravitational potential is
locally homogeneous. For galaxies, this Newtonian approximation
provides a good description giving an exact analytic solution within
the Newtonian limit. The following density profile represents halos in
condensed state or halos in a combination of exited states, $\rho =
\rho_0\sin^2(kr)/(kr)^2$, where $\rho_0$ and $k$ are fitting
parameters. Efforts to include the baryonic influence on SFDM halos
are necessary to have accurate comparisons with observations however
they are still in very early stages of understanding
\citep{Gonzalez13}.

An important feature of SFDM density profiles is the presence of
wiggles, characteristic oscillations of scalar field configurations in
excited states. This could result in some differences in the
gravitational potential for a scalar field halo compared with the 
$\Lambda$CDM profile, possibly imprinting a signature in the stellar 
kinematics of the galactic stellar halo.

The aim of this paper is to generalize the results obtained in paper
I, where we only employed triaxial and spherical NFW halos. Now we
carry out numerical simulations with other galactic components as a
Miyamoto-Nagai disk and a different dark matter density profile, and
other gravity theory (MOND disk). Finally, for this work we also
generalize the initial conditions used in paper I.

This paper is organized as follows. In Section \ref{model} the 3-D
galactic potentials used to compute orbits are briefly described. In
Section \ref{simulations} we introduce our numerical simulations,
techniques and a strategy aimed to efficiently explore the stellar
phase space accessible to a hypothetical observer, we also present the
results of our numerical simulations. In Section \ref{generalizedIC} we
generalize our initial conditions scheme. Finally, in Section
\ref{conclusions} we present a discussion of our results and our
conclusions.

\section{The Potential Models}\label{model} 

In Paper I we assumed a steady NFW \citep{Navarro1996} triaxial dark
matter halo as a proof of concept that a triaxial halo develops and
preserves abundant structure in the stellar phase space because of the
resonant orbital structure. It is worth mentioning that this is
applicable also to other non-spherical halo profiles (oblate,
prolate). With that study we were able to produce clear features in
the velocity space (i.e. halo moving groups), but we did not include
any other galactic components or halo representations other than a
galactic triaxial NFW profile halo.

In this paper, with the purpose of testing the generality of the
results in Paper I we have extended our studies including now the
effect of a disk and a very different triaxial halo potential. We have
also explored a disk that responds to a modified law of gravity
(MOND).

{\bf Our results are applicable regardless the specific values of the
  gravitational potential parameters, they only depend on the
  triaxiality, therefore the kinematic signature is expected either in
  the Milky Way or other galaxies that may have non-spherical
  halos. As the best known case is the Milky Way, and the one that
  nearest in the future will have stellar kinematics observations
  available, we will consider it as our testbed system. However we
  emphasize that the study does not need a detailed model of the Milky
  Way Galaxy.}

\subsection{Halo Potentials} 
We have implemented two intrinsically different potential halos, one
as in Paper I, produced by a NFW density profile \citep{Penarrubia2009}

\begin{eqnarray}
  \Phi(x,y,z)=2\pi G abc\rho_0 r_s^2 \hspace{3.5cm} \nonumber \\ 
\hspace{0.3cm} \times \int_0^\infty \frac
       {s(\tau)}{r_s+s(\tau)} \frac{d\tau } {\sqrt
         {(a^2+\tau)(b^2+\tau)(c^2+\tau)}},
\label{phieq}
\end{eqnarray}
where $\rho_0$ is the characteristic halo density, the dimensionless
quantities $a$, $b$ and $c$ are the three main axes, and $r_s$ is the
radial scale. The triaxiality effect is accounted by using elliptical
coordinates, where,

\begin{equation}
  s(\tau)=\frac{x^2} {a^2+\tau}+\frac{y^2} {b^2+\tau}+\frac{z^2} {c^2+\tau}.
\label{elipeq}
\end{equation}
For the NFW halo the density profile depends on the free parameters
$r_s$, $\rho_0$ and the axis ratios $a, b$ and $c$. As in Paper I, the
rotation curve is used as one of the primary observational constraints
for the parameters, so we kept $r_s=8.5\kpc$, and $\rho_0= 0.056
M\odot$ $pc^{-3}$, obtained assuming a maximum rotation velocity of
$220 \kms$.

For the second halo we used the density profile of a scalar field
configuration \citep{Robles2013} and included the triaxiality using
the formulae of \citet{Chandrasekar1969}, that leads to the next
gravitational potential

\begin{eqnarray}
  \Phi(x,y,z)= \hspace{3.5cm} \nonumber \\ 
\hspace{0.3cm} \times \frac{\pi Gabc\rho_0}{k^2}\int_0^{\infty} \frac{ln(s(\tau)) - Ci(2ks(\tau))}{\sqrt{a^2+\tau}\sqrt{b^2+\tau}\sqrt{c^2+\tau}}d\tau, 
\label{eq:tripotential}
\end{eqnarray}
where $a$, $b$, and $c$ are the three main axes and $Ci$ is the cosine
integral function.

For the scalar field halo we chose the parameters $k=0.123\kpc^{-1}$
and $\rho_0=0.0196$ $M_\odot$ $pc^{-3}$ to guarantee a maximum
rotation velocity of $220$ $\kms$. \textbf{It is worth mention here
  that, with a correct selection of the parameters, the SFDM model has
  proved to be in a good agreement with observational data at galactic
  scales, particularly with rotation curves. And for several cases, a
  SFDM halo fits even better data than a NFW halo (see
  e.g. figs. 5,6,7 and 8 of \citet{MartinezMedina2014}).}

\subsection{Disk Potential}
With the halo potential, we have included a potential that simulates
the Galactic disk. We considered two cases, a Miyamoto-Nagai and a
Kuzmin model for a modified gravity case. With these models we
performed numerical simulations in order to explore the influence of
the disk on the orbital structure of the Galaxy.

\subsubsection{Miyamoto-Nagai potential}\label{M-N}
A commonly used model for the Galactic disk is the one proposed by
\citet{MN75},

\begin{equation}
\Phi(R,z) = -{GM \over \sqrt{R^2+(a+\sqrt{z^2+b^2})^2}}.
\label{MiyNag}
\end{equation}
This potential has three free parameters, $a$ and $b$ that represent
the radial and vertical length-scale respectively, and $M$ the total
mass of the disk. For the case of the Milky Way we adopted $a=5.3178$
\kpc, $b= 0.25$ $\kpc$, and $M= 8.56\times 10^{10}$ M$\odot$
\citep{Allen91}, values that are in good agreement with observations.

\subsubsection{MOND Kuzmin disk potential}
With the purpose of searching for a difference in the orbital
structure (moving groups) induced by using the typical models for the
Galactic potential, that assume the existence of dark matter massive
triaxial halos and modified newtonian dynamical models for the
gravity, we have constructed a MOND (Modified Newtonian Dynamics)
galactic disk as proposed by \citet{M1983}, as an alternative to the
dark matter halo.

Unlike newtonian gravity, the MOND version of Poisson equation is
non-linear and extremely difficult to solve. However, for some mass
distributions, it is possible to find the gravitational potential and
a Kuzmin disk is one of these cases. The advantage of employing a
Kuzmin disk is that the MOND gravitational field can be obtained from
the newtonian field \citep{Read2005},

\begin{equation}
g = g_N \sqrt{{1+(1+4a_0^2/g_N^2)^{1/2}}\over{2}},
\label{gravMOND}
\end{equation}

where $a_0$ is the MOND constant, $g_N$ the newtonian gravitational
field generated by the Kuzmin disk, and $g$ is the corresponding MOND
gravitational field.

Consequently, since the Miyamoto-Nagai potential reduces to a Kuzmin
potential when b=0, that corresponds to the case of a completely flat
disk. The potential produced in this case is

\begin{equation}
\Phi(R,z) = -{GM \over \sqrt{R^2+(a+|z|)^2}}.
\label{kuzmin}
\end{equation}
Thus $g_N$ is calculated first from the Kuzmin potential, and then we
obtain $g$ with equation (\ref{gravMOND}). Once we have $g$, we carry
out the integration.

In the next section, we explore the effect of the halo and the disk
potential in the resulting orbital structure.

\section{Numerical Test Particle Simulations}\label{simulations} 

Our physical system consists of test particles moving under the
influence of a fixed gravitational potential which is the
superposition of different galactic components as described in the
sections above.  The temporal evolution of the system is solved using
the Bulirsh-Stoer method \citep{Press1992}, that works best when
following motion in smooth gravitational fields. Using this tool we
follow, for each particle, the evolution of $x(t)$, $y(t)$ and $z(t)$
by integrating the equations of motion numerically.  As shown in paper
I, the orbital analysis is performed in the velocity space focusing in
the kinematic projections $v_x$ - $v_z$ by means of an orbital
classification. For this purpose we apply the spectral method by
\citet{Carpintero1998} that automatically classifies an orbit,
distinguishing it between a regular and an irregular one. It can also
identify loop, box, and other resonant orbits, as well as high order
resonances.

In our previous work we already saw that for a triaxial halo,
resonances trigger kinematic stellar structure in the case of a single
accretion event and also when the potential is populated in a
stochastic way trying to mimic the stellar halo assembly history.

In order to generalize the results seen in paper I, we follow two
approaches to generate the initial set ups.  The first one is the one
already used in the previous paper, where we choose in advance, the
observation point. The particles are placed randomly around this point
with velocities ranging between zero and the local escape
velocity. This set up will mimic the contribution of many accretion
events with different orientations and energies but keeping only
bounded orbits.

This method has proved to be efficient to explore the available phase
space, as it will be confirmed in the next sections when using more
general initial conditions.  This is because a star that falls in a
resonance will return to the neighbourhood of the observation point,
therefore, we are focusing on orbits that the artificial observer will
detect. The process produce persisting kinematic groups at the
velocities around resonances because irregular and open orbits will
almost never return. In contrast a non-resonant region will present
nearly evenly distributed types of orbits as a result of phase mixing,
and also non-kinematic groups after some mixing time scales. This
method increases our particle statistics without the use of high
computational times.

\subsection{Triaxial halo with disk}
\label{Halo-disk} 

\begin{figure}
\includegraphics[width=0.6\textwidth]{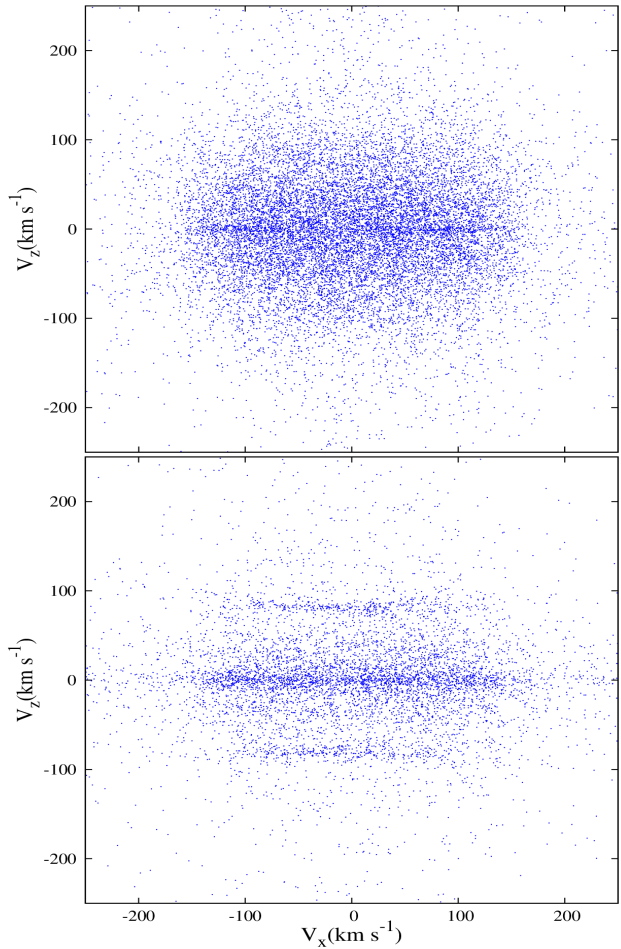}
\caption{Stellar kinematic distribution in the velocity space, as seen
  by an observer at $15 \kpc$ from the halo center, for a spherical
  NFW halo + Miyamoto-Nagai disk (top panel), and for a triaxial NFW
  halo + Miyamoto-Nagai disk (bottom panel).}
\label{fig_haloesfvstriax}
\end{figure}

The observation neighbourhood is defined as a sphere of 1 kpc of
radius and centered at the given observation point, in this case ($x =
15$ kpc, $y = 0$, $z=0$). We set $2\times 10^6$ stars randomly
distributed inside this sphere with velocities that range between zero
and the escape velocity. Because we are focusing in the stellar halo,
stars generated with $v_z \approx 0$ ($v_z/v_x < \tan 10\deg$) are
removed from the initial conditions. These stars will move close to
the X-Y plane, remaining in this plane all along the simulation and
will always have $v_z$ close to zero. If we have populated the whole
halo this orbits concentration would be compensated with neighbour
orbits thrown out from different galaxy positions. This orbits would
appear in the velocity plane as an horizontal band but they do not
form a kinematic group, as we verified it with the spectral orbit
classifier \citep{Carpintero1998}. These particles are an artifact,
already explained in a previous letter \citep{Rojas2012} and should
not be confused with the ones induced by the presence of the disk (see
Fig. \ref{fig_haloesfvstriax}). The system evolves for some 12
Gigayears and at the end of this integration time, we studied the
stellar kinematic distribution. This integration time is long enough
to be comparable to the Milky Way halo evolution
\citep{Kalirai2012}. For our purpose we consider the stars that are
found after the simulation inside the ``artificial observer
neighbourhood''. In Figure \ref{fig_haloesfvstriax} we show the
kinematic distribution of stars at the end of the simulation for two
different cases as seen by an observer at $15 \kpc$ from the halo
center. The upper panel of Figure \ref{fig_haloesfvstriax} corresponds
to a spherical NFW halo $(a=1.0, b=1.0, c=1.0)$, the bottom panel
corresponds to a triaxial NFW halo, where in order to generalize the
results of paper I, we adopt the same values for the axis ratios
$(a=1.47, b=1.22, c=0.98)$, consistent with cosmological simulations
(\citet{Jing2002,Vera2011}).

Both cases contain a Miyamoto-Nagai disk as described in section
\ref{M-N}.  The Miyamoto-Nagai disk is axisymmetric, which is not
necessarily true in the presence of a triaxial halo, since the
potential of the latter could deform the disk.  However, this does not
affect the results because an elliptical disk aligned with the major
axis of the halo only reinforce the kinematic signature of
triaxiality. Note that the only structure that appears in the first
case is a band close to $v_z=0$, corresponding to the stars moving in
the X-Y plane. As explained in Paper I, although these orbits would
appear in the velocity plane as a horizontal band, they do not form a
kinematic group. In the second one we can observe symmetrical bands at
$v_z \sim \pm80$ ${\rm kms^{-1}}$. This figure shows a clear
difference between both cases. The symmetrical bands, associated to
resonant orbits, would be evidence for the presence of a triaxial
halo.

\subsubsection{Changing the axes.}

The previous axes choice for the triaxial dark matter halo allows us
to compare with the results of paper I.

Now {\bf we change the axes ratios (a = 1.47, b = 0.97, c = 0.48)},
and orientate the disk and the dark matter halo according to the halo
model that fits better the Sagittarius stream \citep{Law2009}. In this
way the minor and major axes lie in the galactic plane with the
intermediate axis perpendicular to the disk {\bf and with the new
  values af the axes we obtain}, $b_p/a_p=0.83$ and $c_p/a_p=0.67$,
corresponding to the axis ratios of the potential.

\begin{figure}
\includegraphics[width=0.5\textwidth]{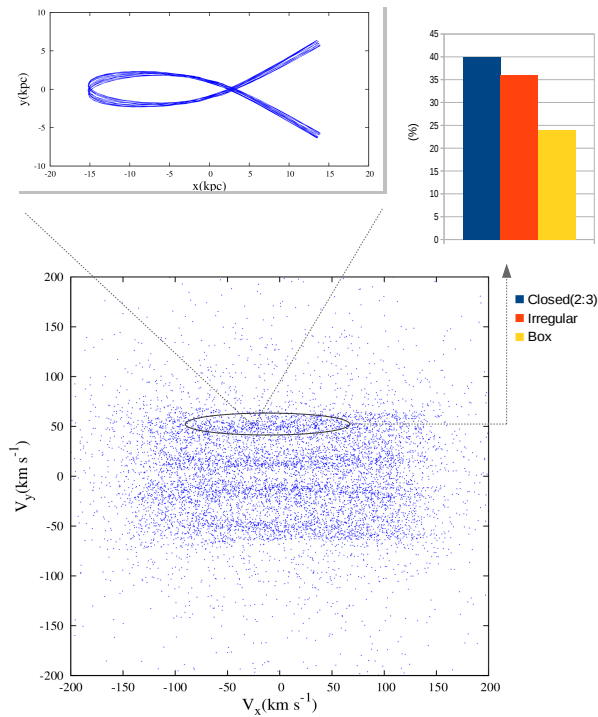}
\caption{$V_x$-$V_y$ projection of the velocity space as seen by a
  hypothetical observer placed at $15 \kpc$ from the center in the
  case of a NFW triaxial halo + Miyamoto-Nagai disk with the axes
  orientation taken from \citet{Law2009}.  The histogram shows the
  result of applying a spectral orbital classifier were we see that
  the regions more densely populated correspond to resonances. This
  supports the interpretation that the kinematic structure has a
  resonant origin. The upper inset shows the orbit of a single
  particle inside one of these resonant regions.}
\label{discohalolaw}
\end{figure}

Figure \ref{discohalolaw} shows the result of this experiment, with an
orientation of the axes different from the treated above but motivated
by observational constraints. The track of the halo triaxiality is
even more visible for this particular orientation as now four
horizontal bands develop in the kinematic stellar structure.  Applying
the stellar orbit classifier to one of this bands, that with $V_y
\approx 50 \kms$, we can establish a resonant origin for these
structures.

Comparing Figures \ref{fig_haloesfvstriax} (bottom panel) and
\ref{discohalolaw}, the differences in the kinematics induced by the
two configurations is noticeable. Both, however contain resonant
orbits in spite the different orientations of the halo relative to the
disk.

\subsection{Kuzmin disk with MOND}\label{Mond}

In this section we describe the numerical simulations carried out
considering a Kuzmin disk within MOND theory \citep{M1983} as an
alternative to dark matter.

In this theory, dark matter does not exist at all. Instead it is
proposed that Newtonian dynamics should be modified to explain
observations. In the MOND version, gravity law has a more general form,

\begin{equation}
\mu({g \over a_0})g = {GM \over r^2},
\label{MOND law}
\end{equation}
where $a_0$ is a universal constant that determines the transition
between the regime of strong and weak field and its value is
approximately $1.2 \times 10^{-10}$ m s$^{-2}$. The function $\mu(x)$ is
not determined in this theory, but it must satisfy the condition
$\mu(x) \approx 1$ when $x \gg 1$ and $\mu(x) \approx x$ when $x \ll
1$.

In the outer part of the galaxies, the weak field regime is valid and
$ \mu(x) \approx x$. This leads to a rotation velocity $v =
{(GMa_0)}^{1/4}$, therefore, the rotation velocity in the outer part
of galaxies is independent of $r$. Thus MOND can explain the
flattening of the rotation curve at great distances from the center of
the galaxies without invoking the existence of dark matter. This is
its main achievement and one of the the reasons it was born.

However MOND also has several problems, especially to correctly
describe the dynamics of galaxy clusters, for example
\citep{Sanders1999,Clowe2004}. Nevertheless, despite that dark matter
is the most accepted theory, MOND (or modifications of it), has not
been totally ruled out.

For this purpose we construct initial conditions in the same way as in
the cases of the triaxial halo and a disk: $2\times 10^6$ stars within
a sphere of $1\kpc$ radius but with the observer neighbourhood
centered at ($x = 15$ \kpc, $y = 0$, $z= 0$), with random velocities
between zero and the escape velocity.

\begin{figure}
\includegraphics[width=0.5\textwidth]{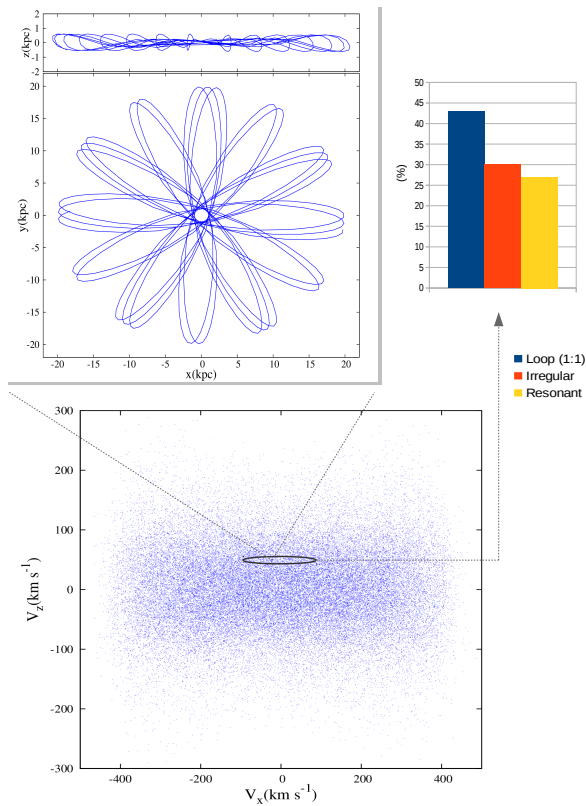}
\caption{$V_x - V_z$ projection of the stellar kinematic distribution
  for a Kuzmin disk within MOND theory. The histogram shows the result
  of the orbit classification. The upper inset shows a loop orbit in
  its projections $x-y$ and $x-z$ for a single particle inside the
  classified region.}
\label{fig_MOND}
\end{figure}

We analyze the resulting velocity space after an integration time of
12 gigayears. Figure \ref{fig_MOND} shows the kinematic structure for
this case. An area of higher density in velocity space is formed, but
without the symmetric bands that characterize the case of a triaxial
halo. 

As shown also in Figure \ref{fig_MOND}, a classification of orbits
give us more information about the kinematic structure, with no
presence of stellar groups and dominated by loop orbits.

\begin{figure}
\includegraphics[width=0.5\textwidth]{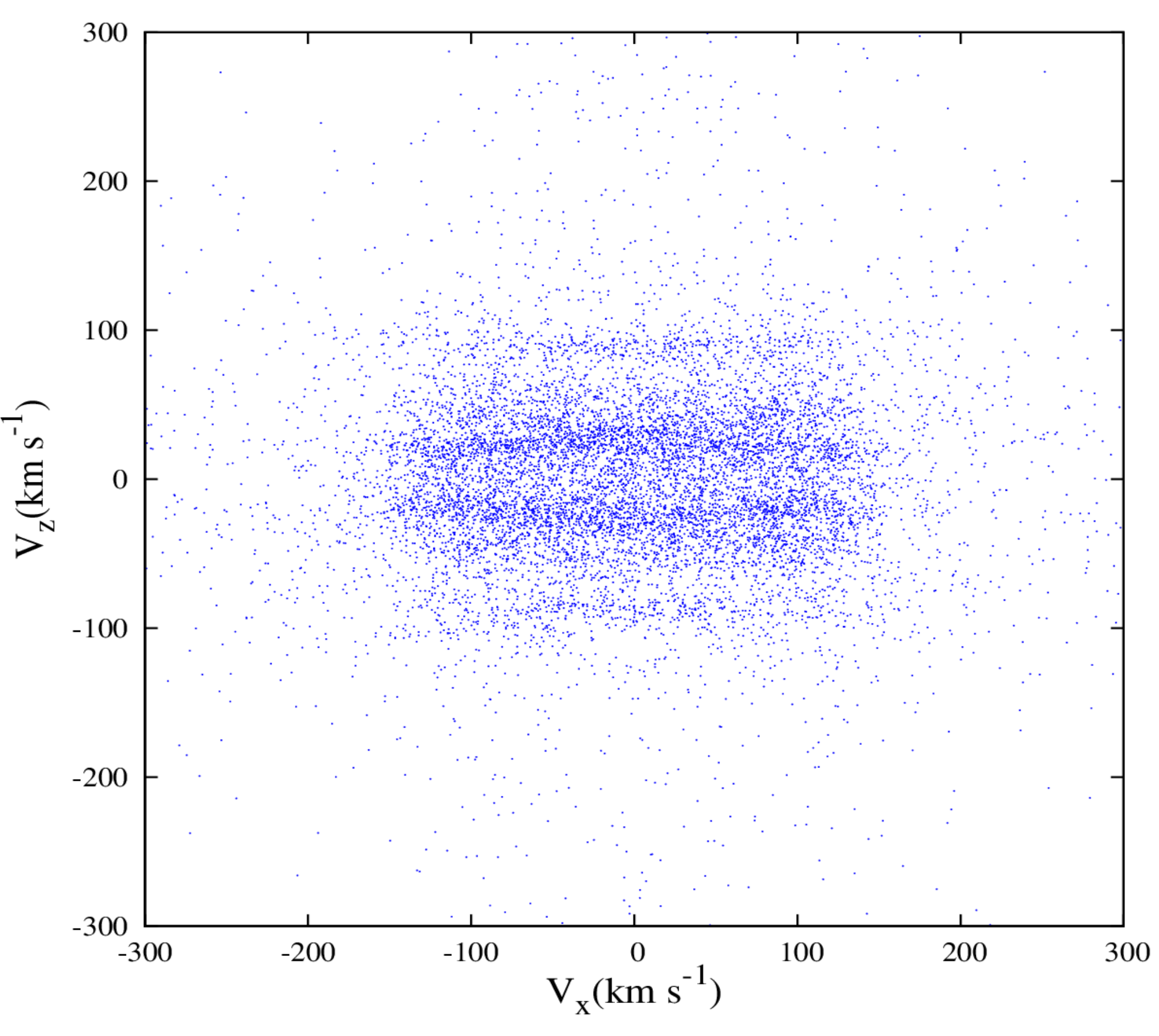}
\caption{$V_x - V_z$ projection of the stellar kinematic distribution
  for a triaxial halo with a Kuzmin disk within Newtonian gravity. The
  presence of kinematic structures is clear in this plot.}
\label{fig_HaloTriaxialKuzmin}
\end{figure}

Although the geometry of the kinematic structure is preserved
regardless of the detailed model of the disk, nevertheless, we decided
to repeat the calculations with a Kuzmin disk within a triaxial halo
with newtonian gravity and found that the structure is very similar to
the one shown in Figure \ref{discohalolaw}, where we used a
Miyamoto-Nagai disk, we show this result in Figure
\ref{fig_HaloTriaxialKuzmin}.

\section{Generalized Initial Conditions}\label{generalizedIC}

As described above the initial conditions set-up, used in the previous
sections and in paper I, provide an observer, with a good kinematic
accuracy, only inside a limited vicinity, and avoiding to populate the
whole halo with particles, ensures a relatively low computing
cost. However, this initial distribution of particles is not isotropic
around the halo center and the question remains about whether this
choice for the initial conditions has a non-negligible effect on the
final kinematic structure for the triaxial dark matter halo employed
in paper I. With this in mind we introduced a more general and
complete initial conditions set-up, different to the described above.

In order to do this, we consider initial conditions with spherical
symmetry around the halo center populating the entire simulated
domain. This is done by increasing the number of particles which are
distributed inside a sphere of $25\kpc$ of radius centered at $(x=0,
y=0, z=0)$. The velocities were selected randomly between zero and
the local escape velocity.

Although computationally expensive, sampling the entire halo and
increasing the number of particles give us a more conclusive mapping
of the gravitational potential to establish its relationship with the
distribution of stars in velocity space. The choice for the initial
conditions ensures that the kinematic stellar structure is not due to
the sampling but due to the physical system.

With these new initial conditions, we now repeat the experiment of
paper I with just a triaxial NFW halo. Finally, a different density
profile for the dark matter halo motivated within the scalar field
dark matter model is also employed. The generation of stellar groups
within these two, very different in nature, triaxial halo potentials
allow us to establish its origin in the triaxiality of the halo rather
than in the dark matter nature.

\subsection{Triaxial NFW halo}\label{NFW}
In this experiment we use the potential of equation (\ref{phieq}) with
parameters $(a=1.47,b=1.22,c=0.98)$ for the axis of the triaxial halo.

We used the generalized initial conditions described above, $16 \times
10^6$ particles populating a sphere of $25\kpc$ of radius centered at
$(x=0, y=0, z=0)$ with velocities selected randomly between zero and
the local escape velocity. The system evolves $12\Gyr$.

\begin{figure}
\includegraphics[width=0.5\textwidth]{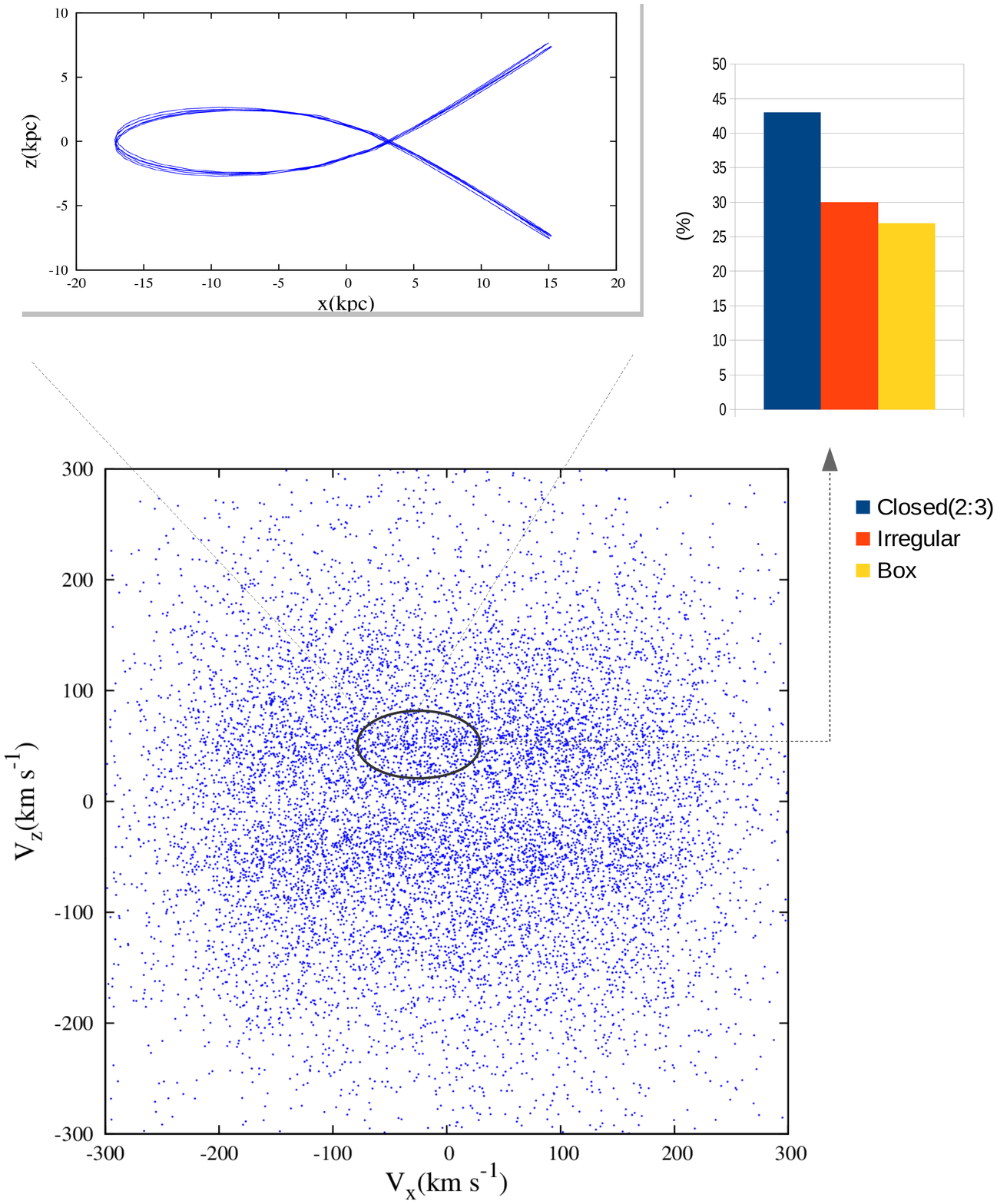}
\caption{$V_x$-$V_z$ projection of the velocity space as seen by a
  hypothetical observer in the case of a NFW triaxial halo. After
  evolving for about 12 Gyr, we found kinematic structure as seen in
  paper I, even with the different initial conditions.  The histograms
  show the result of applying a spectral orbital method were we see
  that the regions more densely populated correspond to
  resonances. This supports the interpretation that the kinematic
  structure has a resonant origin. The upper inset shows the orbit of
  a single particle inside one of these resonant regions.}
\label{HaloTriaxialCompleto}
\end{figure}

Figure \ref{HaloTriaxialCompleto} shows the $V_x$-$V_z$ projection of
the velocity space at the end of the integration time as seen by a
hypothetical observer placed at $15\kpc$ from the halo
center. Overdensities are distinguishable in the kinematic structure,
mostly around $Vz \approx 45 \kms$ and $Vz \approx -45 \kms$. By doing
a classification of orbits one can notice that these overdensities
have a resonant origin and its independent of the initial conditions.

\subsection{Triaxial scalar field dark matter halo}\label{Scalar}
For this triaxial halo (equation \ref{eq:tripotential}) we used the
generalized initial conditions described above, $12 \times 10^6$
particles populating a sphere of $25\kpc$ of radius centered at $(x=0,
y=0, z=0)$, with velocities selected randomly between zero and the
local escape velocity.

\begin{figure}
\includegraphics[width=0.51\textwidth]{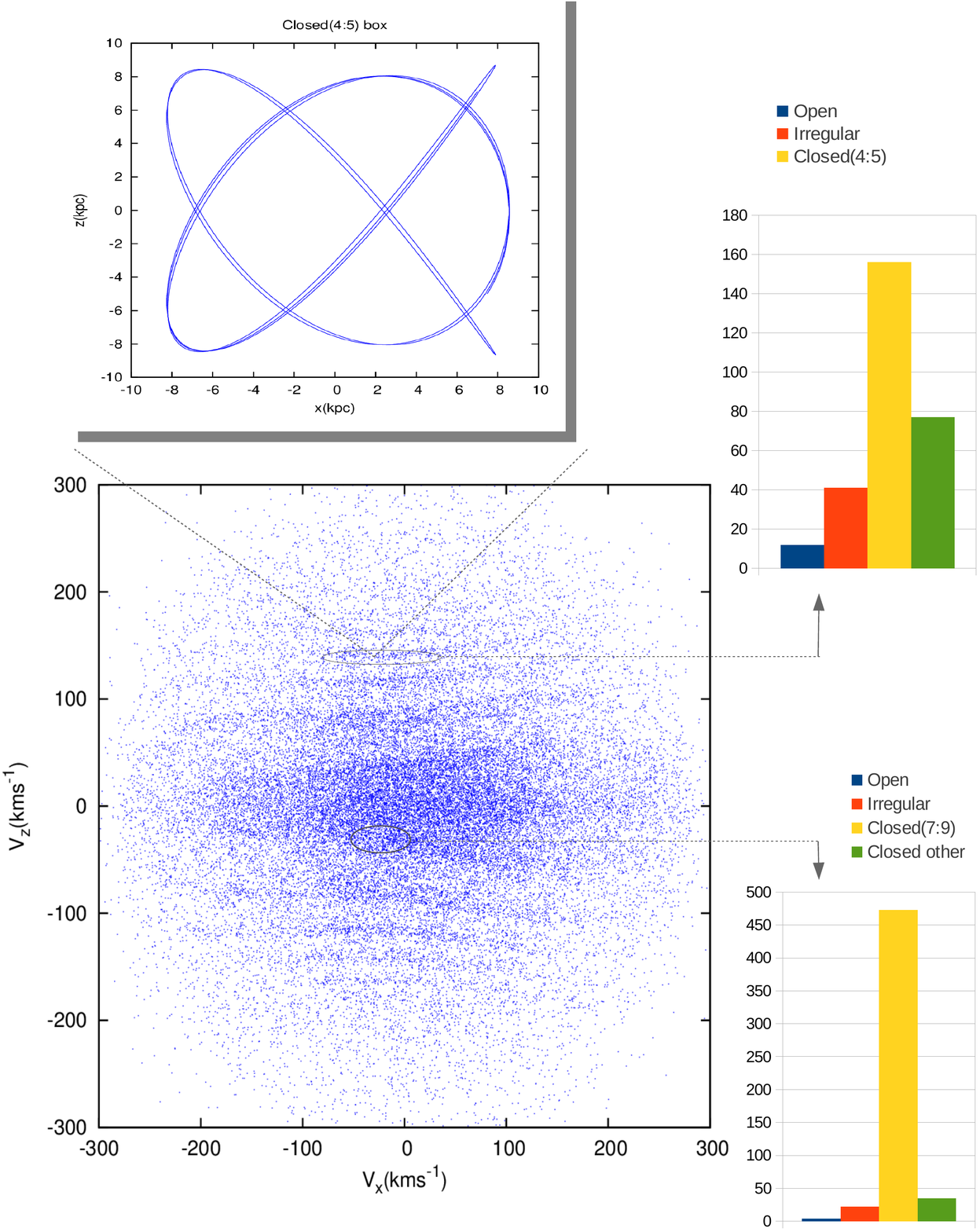}
\caption{$v_x$-$v_z$ projection of the velocity space as seen by a
  hypothetical observer in the case of the scalar field triaxial
  halo. After evolving for about 12 Gyr, we found kinematic structure
  as seen in paper I, even with the randomly selected velocities and
  homogeneous initial particle distribution around the halo
  center. The histograms showed here are the result of applying a
  spectral orbital method, where regions more densely populated,
  correspond to resonances. The upper inset shows the orbit of a
  single particle inside one of these resonant regions.}
\label{fig_region1b}
\end{figure}

We selected $(a=1.47,b=1.22,c=0.98)$ for the triaxial halo axes, and
evolved the system for $12$ \Gyr. Figure \ref{fig_region1b} shows the
$V_x$-$V_z$ projection of the velocity space at the end of the
integration time, as seen by a hypothetical observer placed at $8.5
\kpc$ from the halo center.

In figure \ref{fig_region1b} we found horizontal structure resembling
that seen in our previous experiments. We also performed a simulation
with the spherical case $(a=1, b=1, c=1)$ for comparison, and found
that the velocity space is featureless and homogeneous. This suggests
that the kinematic structure seen in the velocity space of the
triaxial halo corresponds, again, to stars librating in the vicinity
of resonances. We verified this by using the spectral orbital method
of \citet{Carpintero1998}, finding that the most remarkable regions of
the velocity space are dominated by resonant orbits.

\begin{figure}
\includegraphics[width=0.51\textwidth]{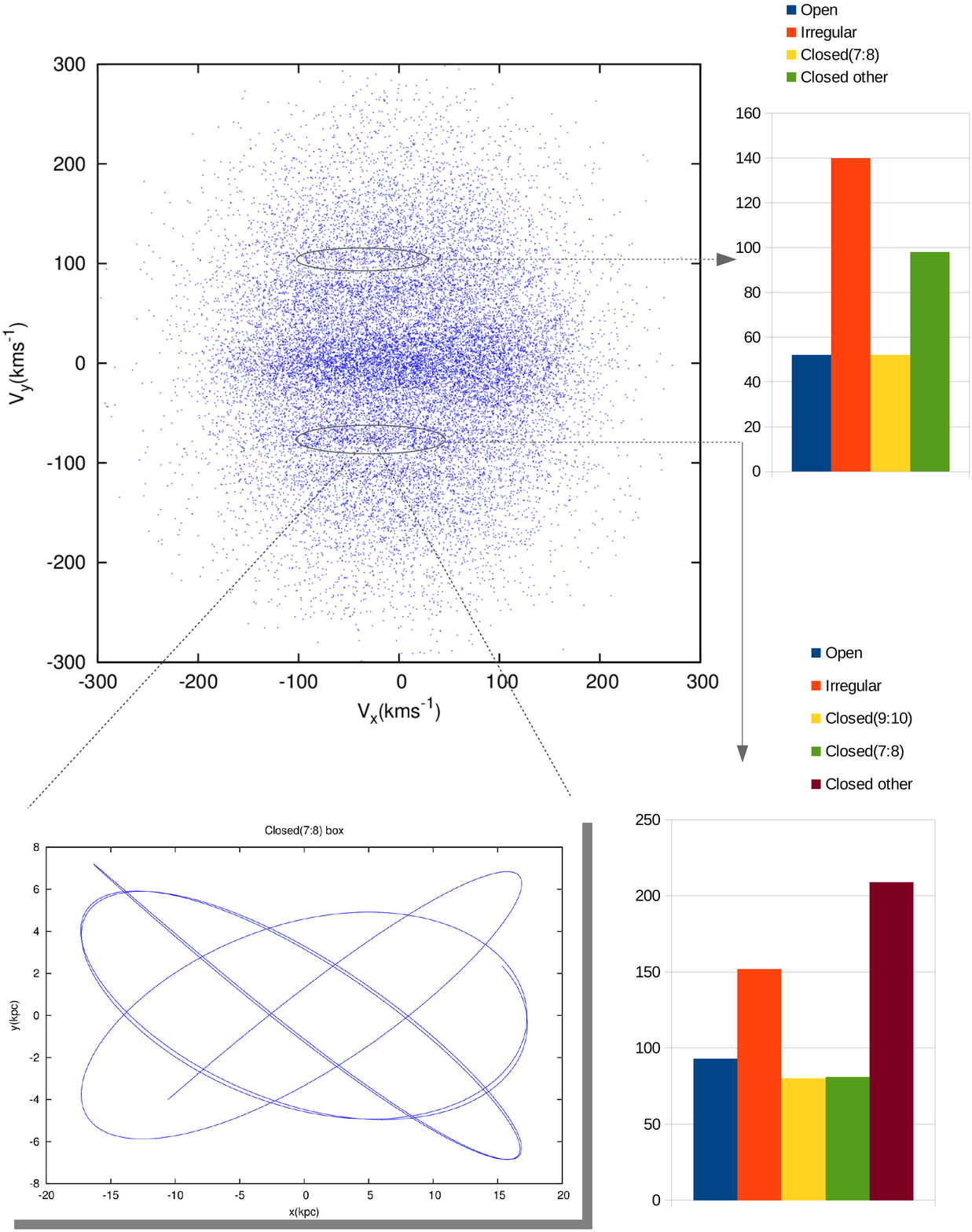}
\caption{$v_x$-$v_y$ projection of the velocity space as seen by a
  hypothetical observer in the case of the scalar field triaxial
  halo. Taken from the same simulation as before but this time the observer 
  is placed at a distance of $20 \kpc$ from the halo center.
  The histograms showed are the result of applying a spectral
  orbital method were we see that the regions more densely populated
  correspond to resonances. This supports the interpretation that
  the kinematic structure has a resonant origin. The bottom inset shows
  the orbit of a single particle inside one of these resonant
  regions.}
\label{fig_region20kpc}
\end{figure}

In figure \ref{fig_region20kpc} we show the $V_x$-$V_y$ projection for
another measure of the data but now moving the hypothetical observer
to a distance of $20 \kpc$ from the halo center.  At this distance we
found again a horizontal pattern as in the figures before which means
that the effect of triaxiality acts across the entire stellar halo.
Again by applying the spectral orbital method we can see that dark
matter halo resonances are able to densely populate the stellar
velocity space with this characteristic horizontal pattern.

\section{Discussion and Conclusions}\label{conclusions}

In this work we extend the orbital study of a stellar system under the
influence of the gravitational potential generated by a steady
triaxial dark matter halo, compared to spherical ones as in Paper I
\citep{Rojas2012}.  We now have added a Milky-Way-like disk to our
study in order to investigate for a way to separate the effects from a
non-spherical halo and a disk.  We have also tested a different type
of dark matter halo profile based on theories of scalar field dark
matter. Finally, we have included a Kuzmin disk assuming a MOND
gravity theory in order to test the consequences on the orbital
behaviour and the possible production of moving groups.

We show that, independently of the nature of dark matter, a
non-spherical shape of the Milky Way dark matter halo, influences
strongly the stellar halo kinematic structure, as seen from the long
lasting resonant features formed in simulations. As it was pointed out
in Paper I, there is an enormous difference with the spherical case,
where no kinematic structure is triggered. We have explored also with
a different halo profile motivated by the scalar field dark matter
model with an important feature, the presence of density wiggles,
characteristic oscillations of scalar field configurations. We
conclude that this scalar field density wiggles are unable to trigger
any moving groups in the spherical case. However in the triaxial case,
this profile produces structures with the same resonant origin as in
the typical NFW triaxial halos.

We also confirm that the structures seen in the velocity space for
triaxial halos, do not depend on the initial conditions, the effect is
still present in spite of the two different initial distributions
strategies of particles used in this paper. The fact that the
kinematic structure is independent from the initial conditions makes
stronger the resonant nature versus an incomplete relaxation of
initial conditions. Our IC´s strategy is motivated by the fact that
there are not selfconsistent phase space distribution for triaxial
halos, it can be argued that incomplete relaxation is producing the
kinematic structure reported in this paper. However the fact that such
structure is produced only in the non-spherical case, and lasts for
longer than the Universe age, and is not altered by the stellar disk
presence makes our conclusions robust.

In this work a disk potential was introduced in order to understand
whether this component is able to erase the orbital structure produced
by a triaxial halo, or if the disk itself could produce moving groups
in the halo that can be erroneously taken as the ones produced by a
triaxial halo. With this study we learnt that the disk influence is
only important close to it, and that it is unable to erase the orbital
structure induced by the halo in all cases we studied. In the same
manner, we studied the effect of the disk on the production of
kinematic structures in the halo, and we found that for an
axisymmetric disk, there are no dominant resonant structures (moving
groups). Although it is known that non-axisymmetric structures (such
as spiral arms and bars) in the disk, may produce important kinematic
structure into galactic disks, their influence diminishes fast out of
the disk plane.

Finally, we have also tested the results comparing with a modified
gravity model (MOND), with a Kuzmin disk. We find in this case no
kinematic structure at all.

In the close future we will be likely able to distinguish, for
instance, satellite remnants groups originated from a single event
from these associated with the dark matter halo shape with the help of
the stellar population data. The great surveys such as Gaia, SDSS,
etc., will allow us the detection of these type of kinematic groups,
which would be a clear indication of the triaxiality (non-sphericity)
of the Milky Way dark matter halo, and will be of great importance for
the galaxy formation theories and the $\Lambda$CDM scenario.

\section*{Acknowledgments}
We thank the anonymous referee of this work for a careful revision
that helped to improve this work. We acknowledge financial support
from UNAM DGAPA-PAPIIT through grants IN114114 and IN117111, and the
General Coordination of Information and Communications Technologies
(CGSTIC) at CINVESTAV for providing HPC resources on the Hybrid
Cluster Supercomputer "Xiuhcoatl".

\end{document}